\documentclass[11pt,twoside]{article}
\usepackage{FUSE2004}
\usepackage{natbib}

\usepackage{epsf}
\usepackage{psfig}
\usepackage{lscape}

\begin{document}
\title{Absorption-Line Spectroscopy of Planetary Nebulae with \emph{FUSE}:
    Probing the Molecular, Atomic, and Ionized Gas}

\author{H.L. Dinerstein and N.C. Sterling}
\affil{Department of Astronomy, 1 University Station C1400, 
   University of Texas, Austin, TX 78712-0259} 

\author{C.W. Bowers}
\affil{NASA/GSFC,  
   Code 681, Greenbelt, MD 20771}

\begin{abstract}

The central stars of planetary nebulae (PNe) are natural
targets for \emph{FUSE} due to their UV brightness.   
The \emph{FUSE} spectra of many PNe show 
absorption features
due to circumstellar material in
species ranging from H$_2$ and neutrals in the photodissociation
region (PDR) to ions resident in the \ion{H}{2} region.
We report results from a program designed to search for
nebular components in 
the H$_2$ Lyman and Werner resonance lines that are responsible
for the fluorescent excitation of H$_2$ in strong FUV radiation fields.
Our failure to detect H$_2$ in absorption in several PNe with
strong near-infrared H$_2$ emission indicates that the molecular 
material has an asymmetrical or clumpy distribution. 
We also detect enrichments in the \emph{s}-process product Ge,
find that Fe is not depleted into dust along at least
one line of sight through a PN, and show that starlight
fluorescence can affect the populations of the excited 
fine-structure levels of \ion{O}{1}.

\end{abstract}

\vspace{-0.2in}

\section{Introduction}

The strong FUV continua of planetary nebula central stars
provide excellent backdrops for observing
absorption lines produced by circumstellar material as
well as foreground interstellar
(IS) clouds. The \emph{FUSE} 
band contains a uniquely broad suite of spectral features
from different phases of material that may be present
in the circumstellar environment, providing
information on physical conditions along a path 
that traverses the entire multi-phase nebular envelope.

An important consideration is the need to resolve
circumstellar from IS features,
for species that may be present in both components.  
We preferentially
selected targets with favorable velocities, such as 
BD+30$^{\circ}$3639, in which   
nebular \ion{Na}{1} absorption is seen at $v_{helio}$ 
$\sim$ --70~km~s$^{-1}$, blueshifted by $\sim$55~km~s$^{-1}$
with respect to the IS components 
(Dinerstein, Sneden, \& Uglum 1995). 

\section{Search for Circumstellar H$_2$}

The primary motivation for our program was to search for
the Lyman and Werner resonance lines that initiate  
the fluorescent excitation of H$_2$. This process 
is one of two chief mechanisms for exciting the near-infrared
(NIR) H$_2$ quadrupole emission lines
seen in many PNe, the other being 
collisional excitation in shock-heated gas 
(e.g.\ Black \& van Dishoeck 1987). 
Although initially the NIR emission in PNe was attributed to
shocks, it has become clear that 
fluorescence plays an important role as well
(Dinerstein et al.\ 1988; Hora, Latter, \& Deutsch 1999). 
\emph{FUSE} directly 
samples the pump lines from the $v$~=~0 levels of the
ground state 
(e.g. McCandliss 2003). Furthermore,   
because of their large oscillator strengths,
the FUV lines are sensitive  to much smaller
column densities of H$_2$ than can be detected
through NIR emission. 

\vspace{-2 mm}

 \begin{table}
  \caption{Results of \emph{FUSE} H$_2$ Search}
    \vspace{.1in}	
    \begin{tabular}{lccc}
     \hline \hline
    \vspace{.05in}	
     {Object} & {IR H$_2$ Emission} & {FUV H$_2$ Absorption} \\
     \hline
     BD+30$^{\circ}$3639 	& yes 	& yes 	\\
     NGC 40 		& yes	& no \\
     NGC 5882 		& unknown & no \\
     NGC 6210 		& no	& no \\
     NGC 6720 		& yes	& no \\
     NGC 7662 		& no	& no \\
     SwSt 1 		& yes	& no \\
     \hline \hline
    \end{tabular}
  \label{sample-table}
 \end{table}

\vspace{-1 mm}

Table 1 summarizes the results for our program.
Although four of the targets
are known sources of
NIR H$_2$ emission, 
only one --
BD+30$^{\circ}$3639 -- shows nebular H$_2$ absorption.
This object also shows excited 
H$_2$ in its \emph{HST}-STIS UV spectrum, which 
contains hundreds of lines from $v$~$\ge$~2 in the ground   
electronic state (Dinerstein \& Bowers 2004; Dinerstein et al,
in preparation). The \emph{FUSE} spectrum
of BD+30$^{\circ}$3639 resembles that 
of M27, the Dumbbell Nebula, which is rife with nebular H$_2$  
(McCandliss et al. 2000). On the other hand, no nebular
H$_2$ is seen toward
the central star of NGC~6720, the Ring Nebula,
where the NIR H$_2$ emission is clumpy and concentrated 
near the bright ring (Speck et al. 2003). 
The \emph{FUSE} results are consistent with 
the H$_2$ residing in an equatorial torus,
since the Ring is seen pole-on,
whereas the Dumbbell is nearly equator-on.

>From the \emph{FUSE} data, we set upper limits of
$N_{J}$(H$_2$)~$\le$~5~$\times$~10$^{13}$~cm$^{-2}$ 
for $J$~=~2 -- 5 on the circumstellar components   
for three PNe with NIR H$_2$
emission: NGC~40, NGC~6720, and SwSt~1. 
These values are  
$\sim$~50~times lower than the actual column densities
in BD+30$^{\circ}$3639, and also 
than the beam-averaged values of $N_J$(H$_2$)
from the NIR emission \emph{in the same PNe.} 
We conclude that the H$_2$ in these PNe has a 
very non-uniform distribution,  
being either globally 
asymmetrical (e.g. in a torus) or concentrated 
in dense clumps.  Only in a few 
cases does the 
line of sight to the central star intercept
circumstellar H$_2$.

\section{Excitation of \ion{O}{1}}

The $^{3}P$ ground term of \ion{O}{1} has  
two excited fine-structure levels 
that give rise to forbidden transitions at 63~$\micron$ 
and 145~$\micron$. 
For $n$~$\ge$~10$^3$~cm$^{-3}$, 
[\ion{O}{1}] 63~$\micron$
is the strongest cooling line from PDRs, the
transition regions between ionized and molecular material. 
Ratios of the [\ion{O}{1}] lines and
[\ion{C}{2}]~158~$\micron$ are diagnostics of
physical conditions and PDR parameters  
(Kaufman et al. 1999).

The populations of the \ion{O}{1} fine-structure levels can
also be measured via UV absorption lines.  
The 1302--1306 \AA\ triplet (observable with \emph{HST})
is often saturated, and the line from the first excited
level, 1304 \AA, is blended with \ion{Si}{2}. 
The \emph{FUSE} band includes other, weaker transitions
which can be used to derive the level populations.  
For example, the \ion{O}{1} triplet near 1040~\AA\ in SwSt~1
yields a value of $N$(\ion{O}{1}**)/$N$(\ion{O}{1}*) which is 
twice the ratio of statistical weights. This is
possible only if a non-collisional process dominates the
excitation. Based on the reported strengths of the optical \ion{O}{1} 
lines   
(De~Marco et al. 2001), we identify this 
mechanism as
fluorescent excitation by
stellar continuum photons (Sterling et al., in preparation).
In SwSt~1 the effect is obvious; 
elsewhere, it could be more insidious, 
elevating the population of the
$^3$P$_0$ level and strengthening [\ion{O}{1}] 
145~$\micron$, thereby leading to an  
overestimate of the gas temperature.

\section{Conditions and Abundances in the Ionized Gas}

The UV spectral region also contains resonance
lines from 
ions in the \ion{H}{2} region.
These can provide diagnostics of physical conditions 
and values for gas-phase abundances. 
We used \emph{FUSE} observations of UV lines 
from the ground and several excited fine-structure levels of 
\ion{Fe}{3}, \ion{Fe}{2}, and \ion{S}{3}  
to derive the gas-phase Fe/S ratio in SwSt~1,
for which we find an essentially solar value
(Sterling et al., in preparation). 
This result suggests that Fe is not depleted into dust
along this particular path through the nebula,
in marked contrast to the results found from the optical and NIR 
emission lines, which 
indicate depletion 
factors of $\ge$~15. The implied inhomogeneity in
the dust-to-gas ratio echos the inhomogeneity 
seen in the H$_2$ in this PN (see \S 2).

\emph{FUSE} also offers access to some species that cannot be
observed in other spectral regions. 
For example, the neutron-capture element Ge ($Z$~=~32)
can be observed via the \ion{Ge}{3}
resonance line at 1088~\AA\ (Sterling, Dinerstein, \& Bowers 2002).
The Ge abundance in PNe can be enhanced by the
dredge-up of 
\emph{s}-processed material during the AGB phase.
We find that Ge/S is enriched by factors of three or more in 
a few PNe observed 
by \emph{FUSE}.

\acknowledgements  We thank the operations and user-support 
staff of the \emph{FUSE} project for all their help in planning
and executing this observing program. Support for this research
was provided by NASA Goddard Space Flight Center through data
analysis grants NAG~5-9239,
5-11597, and 5-12731.

\end{document}